# Vapor-liquid equilibrium predictions of n-alkane/nitrogen mixtures using neural networks

Suman Chakraborty[1*], Yixuan Sun[2*], Guang Lin[2,3], Li Qiao[1]

[1]School of Aeronautics and Astronautics, Purdue University, West Lafayette, IN 47907, USA

[2]School of Mechanical Engineering, Purdue University, West Lafayette, IN 47907, USA

[3]Department of Mathematics, Purdue University, West Lafayette, IN 47907, USA

**Abstract**

Understanding fluid phase behavior in high pressure and high temperature conditions is crucial for developing high-fidelity simulations of chemically reacting flows in liquid-fueled combustion systems. The study of vapor-liquid equilibrium (VLE) curves also forms an integral part of the design and modeling of the control processes in chemical and oil - gas industries. The main objective of this study was to develop data-driven models to predict VLE of Type III binary mixtures involving long-chained n-alkanes and nitrogen. Two data-driven models have been proposed in this study, each of which was competent in estimating VLE for the binary systems of $C_{10}/N_2$ and $C_{12}/N_2$, at pressures ranging up to 50 – 60 MPa. Both the models showed better performance (less average absolute percentage error) in predicting equilibrium pressure of the binary mixtures as compared to the VLE modeled using Peng-Robinson equation of state (PR-EOS). The data-driven models were also able to correctly trace the change in curvature of the vapor phase composition, close to the mixture critical point, at high pressures and temperatures – a feature of the n-alkane/nitrogen system VLE which the PR-EOS model fails to capture.

## 1. Introduction

Phase equilibrium calculations play a crucial role in the study of high-pressure liquid fuel injection processes, involving a number of complex physical processes such as mixing, atomization and evaporation [1–4]. Understanding of phase equilibrium behavior, specifically, vapor-liquid equilibrium (VLE) forms an important premise in the design of distillation processes, widely used in the chemical industry [5]. Even in the oil and gas sector, VLE behavior of nitrogen/alkane systems helps to understand nitrogen solubility, which in turn can be utilized in designing the nitrogen injection systems used for oil recovery [6]. The design and optimization of the control facilities of the aforementioned applications heavily rely on VLE data.

Binary mixtures of n-alkanes/nitrogen, except for methane, follow a Type III phase behavior according to the works of Scott and van Konynenburg [7]. The discontinuous critical line phase behavior exhibited by Type III mixtures poses challenges in its theoretical prediction [8,9]. One part of the critical locus, starts at the critical point of the heavier species and rises up to infinitely high pressures, whereas the other starts at the critical point of the lighter species and ends up meeting a three-phase vapor–liquid–liquid equilibrium (VLLE) coexistence line. Several experimental studies [10–16] have reported the vapor–liquid equilibrium behavior of n-alkane/nitrogen binary systems. However, the experimental database of VLE data for heavy n-alkane/nitrogen mixtures is scanty, with very few studies on heavier alkanes and even fewer in the high pressure and temperature range [6,17].

VLE modeling using equation of state (EOS) is another way to study the phase behavior of n-alkane/nitrogen systems. The Peng–Robison equation of state (PR-EOS) [18] is one of the more popular EOS models for heavier alkanes [19]. However, there is a downside to using the EOS models. Firstly, the equation of states involves empirical parameters, especially for mixing rules which require a binary interaction parameter (BIP) for the accurate modeling of VLE using EOS, which is hard to determine [20]. Some studies proposed a constant BIP for a mixture system by minimizing the difference between experimental and EOS modeled pressures and vapor compositions [6,17]. Fateen et al. [20] proposed a semi-empirical relation for BIP based on Huron-Vidal mixing rules, for over 60 binary mixtures, which also allowed the BIP to exhibit unsymmetrical behavior. Some studies [21,22] have proposed generalized BIP correlations to be used with PR-EOS, making use of limited experimental data in literature [20]. Another issue of VLE modeling using EOS is the iterative nature of the process, which can cause delays for real-time, on the fly calculations in an industrial application [23].

Development in the fields of data-driven (DD) learning and neural networks (NN) offer a fast and computationally efficient alternative to solving VLE problems [23–32]. Neural networks have also been used in estimating other thermodynamic and transport properties of mixtures [33–37]. NNs are capable of capturing the non-linear and complex patterns between a set of parameters, without any prior knowledge of the governing equations or physics of the problem. For a VLE problem, a network can be trained to predict the non-linear mapping between the equilibrium temperature (T), pressure (P) and composition ($x, y$) in liquid and vapor phase [25]. The study by Sharma et al. [25] explored the use of backpropagation algorithm in the VLE prediction of methane/ethane and ammonia/water systems, and found the results to be within $\pm 1\%$ accuracy. Urata et al. [38] used an artificial neural network (ANN) to predict the VLE behavior of binary mixtures consisting of hydrofluoroethers (HFEs) and polar compounds. Mohanty [23,27] also used ANN to study VLE for the binary systems of carbon dioxide/difluoromethane, ethyl caproate, ethyl caprylate, and ethyl caprate, the likes of which are either used as refrigerants or are significant in supercritical extraction. A Least Squares Support Vector Machine (LSSVM) model, with a $R^2$ of 0.9932, was used by Mesbah et al. [39] in predicting the phase behavior of 7 carbon dioxide/hydrocarbon binary mixtures. In a more recent study by Roosta et al. [40], ANN was used to predict the VLE of methane to n-heptane binary mixtures, and proposed closed-form expressions using genetic programming.

Motivated by the above, this study uses data-driven models for predicting the VLE of two mixtures: n-decane/nitrogen ($C_{10}/N_2$) and n-dodecane/nitrogen ($C_{12}/N_2$), at pressures up to 50–60 MPa. The data-driven models formulated as a part of this study have been made with the goals of:

I. To bypass the need for estimating and/or formulating a binary interaction parameter. The process and experimental data required for the same are either non-existent, or hard to acquire
II. A model that can be trained on the minimal experimental data points available, and is still able to capture the vapor-liquid equilibrium behavior at high temperatures and pressures

Keeping the goals in mind, two different data-driven unified models were proposed. The first model (DD-Int) was trained based on an interpolated data set generated from the experimental data and worked for both binary mixtures. The second unified model (DD-Ind-T) has been trained using the data generated from individual sub-models, which were trained on temperature specific (isotherm) VLE data. The equilibrium pressure (P) and vapor phase mole fraction ($y$) predicted using the data-driven models were then compared with the experimental data and PR-EOS modeled data.

## 2. VLE modeling using EOS

In this section, the formulations for VLE modeling using an equation of state will be described. For a mixture to be in vapor-liquid equilibrium, the liquid (L) and vapor (V) phases must satisfy the following conditions:

$$T^L = T^V \quad (2.1)$$

$$P^L = P^V \quad (2.2)$$

$$\hat{f}_i^L = \hat{f}_i^V \quad (2.3)$$

where, $\hat{f}_i^L$ and $\hat{f}_i^V$ are the fugacities of the i-th component of the mixture in the liquid and vapor phase, respectively. Before proceeding, an equation of state must be chosen to arrive at the fugacity formulations. For this study, Peng–Robison equation of state (PR-EOS) [18] has been used. For a pure fluid, PR-EOS can be expressed as the following [18]:

$$P = \frac{RT}{v-b} - \frac{a(T)}{v(v+b) + b(v-b)} \quad (2.4)$$

where, $a$ and $b$ are the two constants of the equation of state, which can be expressed for pure components as:

$$a = 0.45724 \frac{R^2 T_c^2}{P_c} \alpha(T_r, \omega) \quad (2.5)$$

$$b = 0.07780 \frac{RT_c}{P_c} \quad (2.6)$$

$$\alpha(T_r, \omega) = \left[1 + \kappa\left(1 - T_r^{0.5}\right)\right]^2 \quad (2.7)$$

$$\kappa = 0.37464 + 1.54226\,\omega - 0.26992\,\omega^2 \quad (2.8)$$

where, $T_r$ is the reduced temperature defined as $T/T_c$ and $\omega$ is the acentric factor. A modification was proposed to the expression of κ in 1978 for heavier components having acentric factor greater than 0.49 [42].

$$\kappa = 0.379642 + 1.485030\,\omega - 0.164423\,\omega^2 + 0.016666\,\omega^3 \quad (2.9)$$

For extending the equation of state to multi-component systems, the van der Waals one fluid mixing rule can be used to re-calculate the new mixture constants as shown:

$$a = \sum_i \sum_j x_i x_j a_{ij} \quad (2.10)$$

$$b = \sum_i x_i b_i \quad (2.11)$$

$$a_{ij} = (1 - k_{ij})\sqrt{a_i a_j} \quad (2.12)$$

where, $k_{ij}$ is the binary interaction parameter between components $i$ and $j$ of the mixture and satisfies $k_{ij} = k_{ji}$ and $k_{ii} = 0$. Equation 2.4 can be re-written in terms of the compressibility factor ($Z = Pv/RT$) as:

$$Z^3 - (1 - B)Z^2 + (A - 3B^2 - 2B)Z - (AB - B^2 - B^3) = 0 \quad (2.13)$$

where, $A = aP/(RT)^2$ and $B = bP/RT$. Using equation 2.13 to solve for the compressibility factor of a mixture in any phase, either result in yielding all real roots or one real root, along with two other complex conjugates. In case of the former, the largest and the smallest root correspond to the compressibility of the mixture in vapor and liquid phase, respectively. The latter case, however, indicates the presence of a single phase.

Fugacity of a component can be computed using the expression for coefficient of fugacity ($\widehat{\phi}_i$), which can be expressed as:

$$\ln \widehat{\phi}_i = \frac{b_i}{b}(Z - 1) - \ln(Z - B)$$
$$- \frac{A}{2\sqrt{2}B}\left(\frac{2\sum_{j=1}^{N} x_j a_{ij}}{a} - \frac{b_i}{b}\right) \ln \left(\frac{Z + (1 + \sqrt{2})B}{Z + (1 - \sqrt{2})B}\right) \quad (2.14)$$

where, $x_j$ is the mole fraction of the j-th component in a mixture of $N$ components. In this study, binary systems ($N = 2$) of $C_{10}/N_2$ and $C_{12}/N_2$ are modeled. $N_2$ will be referred to as component – 1 ($x_1$, $y_1$) and the n-alkanes as component – 2 ($x_2$, $y_2$).

# 3. Neural Network Theory

Inspired by nature, neural networks, especially deep neural networks have been widely used in solving data-driven problems in recent years. The universal approximation theorem [43] states that it is possible to represent any function using a simple feed-forward neural network with only a single hidden layer. Neural networks have been trained to model a wide range of functions in an input-output paired supervised way as they are powerful function approximators. The development of deep learning techniques overcame the problem that the number of nodes in a single-layered neural network can be infeasibly large when approximating complex functions. Gradient-based optimization methods, backpropagation [44], and high-performance computing made it possible to efficiently train very deep neural networks.

Deep neural networks have been making major advances in solving various engineering problems, such as mechanical properties prediction of porous media [45,46], outage prediction and deriving generation dispatch in power grids [47,48], and material temperature measurement and crack detection based on infrared thermography [49,50]. In this work, we mainly focus on a basic form of neural network - feedforward fully connected networks to inspect their ability to model the VLE of Type III binary mixtures, involving long-chain n-alkanes, at different temperatures with limited experimental data. Feedforward networks comprise of an input layer, several hidden layers and an output layer. In general, the feedforward fully connected networks take the input information at the input layer and then pass it to each node in the subsequent layer, which can be described as:

$$\boldsymbol{Y} = \boldsymbol{X}\boldsymbol{W} + \boldsymbol{b} \quad (3.1)$$

where $\boldsymbol{X} \in \mathbb{R}^{n \times m}$ is the input matrix containing $n$ data points and $m$ features; $\boldsymbol{W} \in \mathbb{R}^{m \times h}$ is the weight matrix, in which $h$ is the number of nodes in the subsequent layer; $\boldsymbol{b} \in \mathbb{R}^{h}$ is the bias vector. At each node in the subsequent layer, an activation operation is implemented. It introduces non-linearity into the network, enabling it to capture the non-linear relations between the input information and the output. The final output from the subsequent layer is:

$$\boldsymbol{O} = a(\boldsymbol{Y}) = a(\boldsymbol{X}\boldsymbol{W} + \boldsymbol{b}) \quad (3.2)$$

where $a(\cdot)$ is the activation function, which is usually a non-linear and differentiable function. The output of the current layer becomes the input to the next subsequent layer and the same process is repeated until the output layer. The value from the network's output layer can be compared with the real value, given the input, using a suitable loss function. Gradient-based optimization methods with backpropagation can help search for the optimal weights that minimize the loss.

## 4. Learning Scheme

In this section, we introduce the process of data augmentation, the choice and implementation of deep learning models, and the evaluation metrics used for measuring model performance. Three different neural network architectures were used in this work.

## 4.1 Data Augmentation

Training an effective deep neural network requires adequate data, especially when the network is complex. Due to the fact that the experimental data for each temperature only contained about 15 – 27 data points, augmenting the experimental data for model training was necessary. Two augmentation methods, linear interpolation and data generation from individually trained temperature-dependent models were adopted in this work to enlarge the training dataset size. The augmented data were used to train a model that took temperature (T), liquid phase composition of the mixture ($x_1$, $x_2$) as inputs, and then output equilibrium pressure (P) and vapor composition of $N_2$ ($y_1$).

### 4.1.1 Linear Interpolation

Despite being simple, interpolating samples in feature space is a useful way to augment the dataset, which is able to effectively reduce model overfitting. Consider an experimental dataset having $N_{exp}$ points, thus having ($N_{exp} - 1$) data slots. The goal is to generate $N_{inter}$ new points using interpolation, such that each of the experimental data slots will house $N_{new-eachslot} = \left(N_{inter} / N_{exp} - 1\right)$ points. For a particular isotherm, for every two experimental data points of x, y and P, $N_{new-eachslot}$ number of new points were inserted. The new points were generated using section formula as follows:

$$x_{new} = \frac{(fact1 \times x_{exp,i+1}) + (fact2 \times x_{exp,i})}{fact1 + fact2} \qquad (4.1)$$

where, $x_{exp,i}$ and $x_{exp,i+1}$ are two consecutive experimental data points between which the new points were generated. The new point ($x_{new}$) divided the line segment joining the two experimental points in the ratio of $fact1:fact2$, where both $fact1$ and $fact2$ were positive integers and satisfied eqn. 4.2. $fact1$ could take values between $(1 + N_{new-eachslot})$ and the corresponding value of $fact2$.

$$fact1 + fact2 = N_{new-eachslot} + 1 \qquad (4.2)$$

### 4.1.2 Data Generation from Temperature – specific Models

Another approach to data augmentation was to generate data from individual models trained on specific temperatures. The experimental data contained a few samples at various temperatures. While the size of dataset was not enough for a model that took the temperature as one of the features, training models conditioned on temperature led to good performance. That is, for each temperature in the experimental dataset, a neural network was trained only on the VLE profile of the mixture for that isotherm. And then the trained models were used to generate new data points by taking randomly sampled $x_1$ and $x_2$ values. Finally, the new data generated from those individual models conditioned on temperatures were used to train a single model that takes $x_1$, $x_2$ as well as the T as the input to predict $y_1$ and P.

### 4.2 Network Architecture

Based on the nature of the tabular format of the data, all three aforementioned architectures were fully connected feed-forward neural networks. A fully connected network comprises of an input layer, an output layer, and several hidden layers, with nodes connected to each other between adjacent layers.

### 4.2.1 Temperature – specific, Single Composition Models

To address the dataset size problem, individual neural network models were adopted to generate VLE profile data conditioned on temperatures. In each individual network, the input layer had two nodes, taking $x_1$ and $x_2$, respectively, and the output layer contained 2 nodes, giving out $y_1$ and P. The architecture of the individual models is shown in Fig. 1.

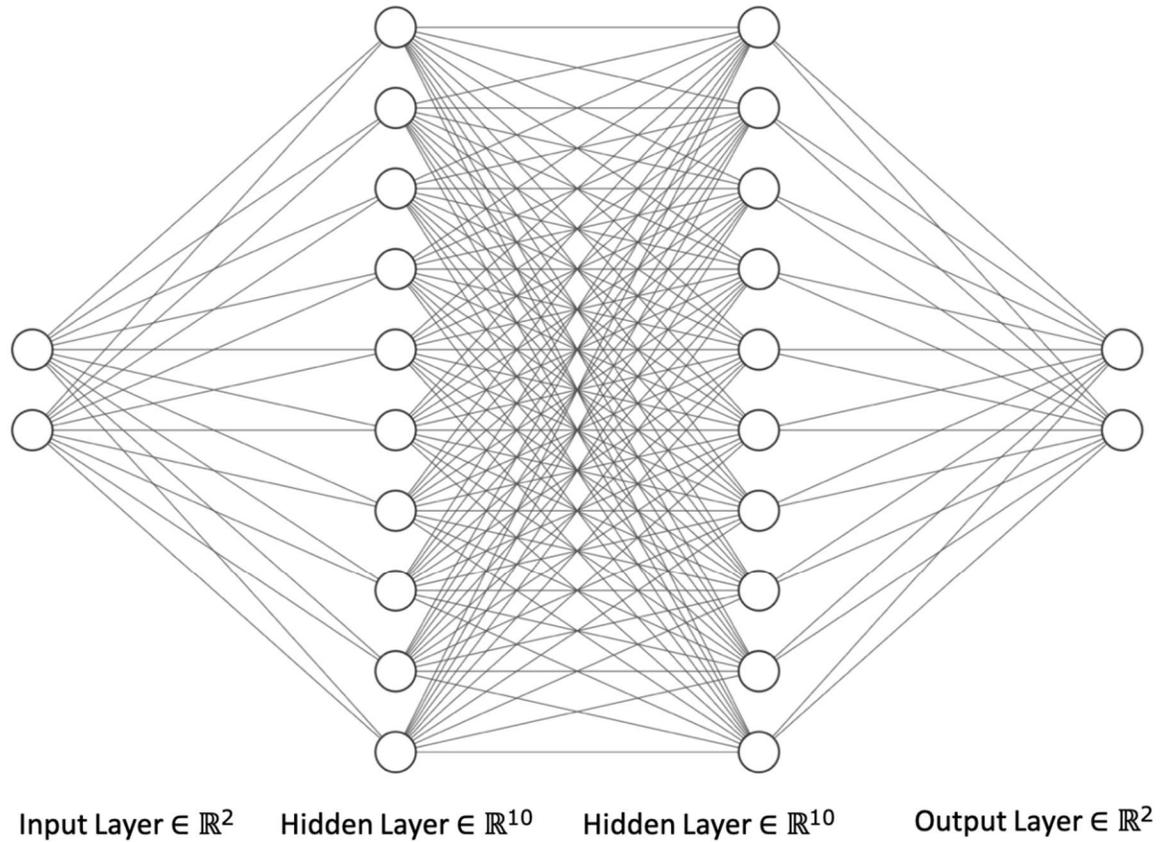

Input Layer ∈ $\mathbb{R}^2$    Hidden Layer ∈ $\mathbb{R}^{10}$    Hidden Layer ∈ $\mathbb{R}^{10}$    Output Layer ∈ $\mathbb{R}^2$

Fig. 1: Temperature – specific network architecture

A rectified linear unit (ReLU) activation function was placed at each node in the hidden layers, which introduces non-linearity to the network. From the last hidden layer to the output layer, a linear mapping was introduced so that the output values were not constrained by any activation to serve the purpose of regression.

### 4.2.2 Composition – specific Models

The augmented data were fed into a model that took T, $x_1$, and $x_2$ values of a specific binary mixture as the input to predict the corresponding $y_1$ and P. In the composition-specific model, there were 3 hidden layers with 10 nodes in each layer. While it had a similar structure with respect to the temperature-specific, single composition model, on the node for $y_1$ value in the output layer, a sigmoid activation was placed to constrain the output to [0, 1], shown in Fig. 2. The same architecture was used to train models for both the binary mixtures of $C_{10}/N_2$ and $C_{12}/N_2$.

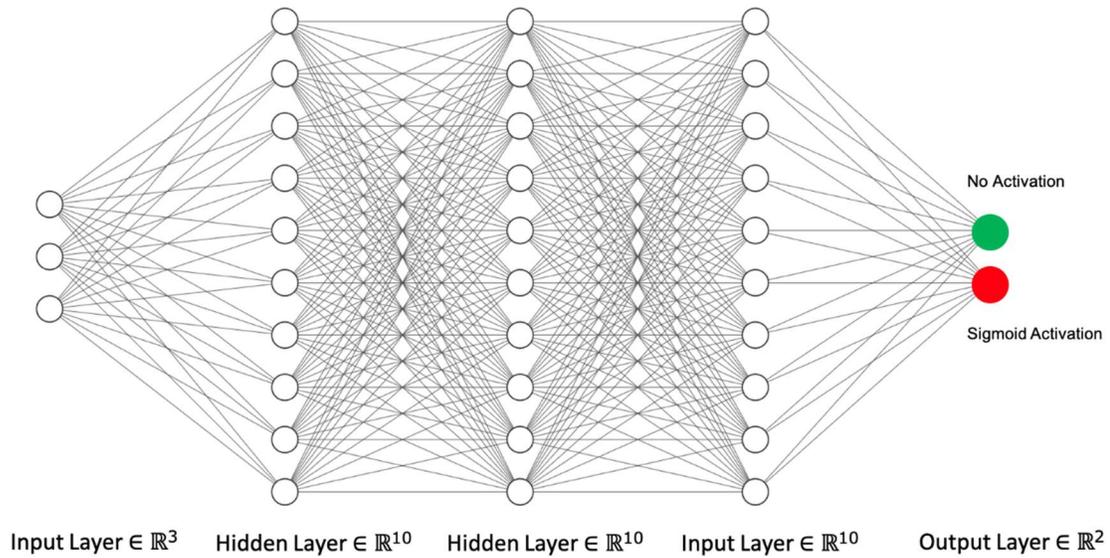

Fig. 2: Composition – specific model. The input layer takes $x_1$, $x_2$ values of a binary mixture, as well as the temperature. On the node for $y_1$ in the output layer, a sigmoid activation is used for constraining the output value.

### 4.2.3 Combined Composition Model

In addition, we explored deep neural networks' ability to model the VLE of more than one binary system. The combined composition model took $x_1$, $x_2$, T, and a model identifier that indicates whether it was a $C_{10}/N_2$ or $C_{12}/N_2$ system, as the input and then output the corresponding $y_1$ and P. This network had 4 nodes in the input layer and 4 hidden layers with 10 nodes in each. Similarly, a sigmoid activation was employed at the node giving out $y_1$ value in the output layer, shown in Fig 3.

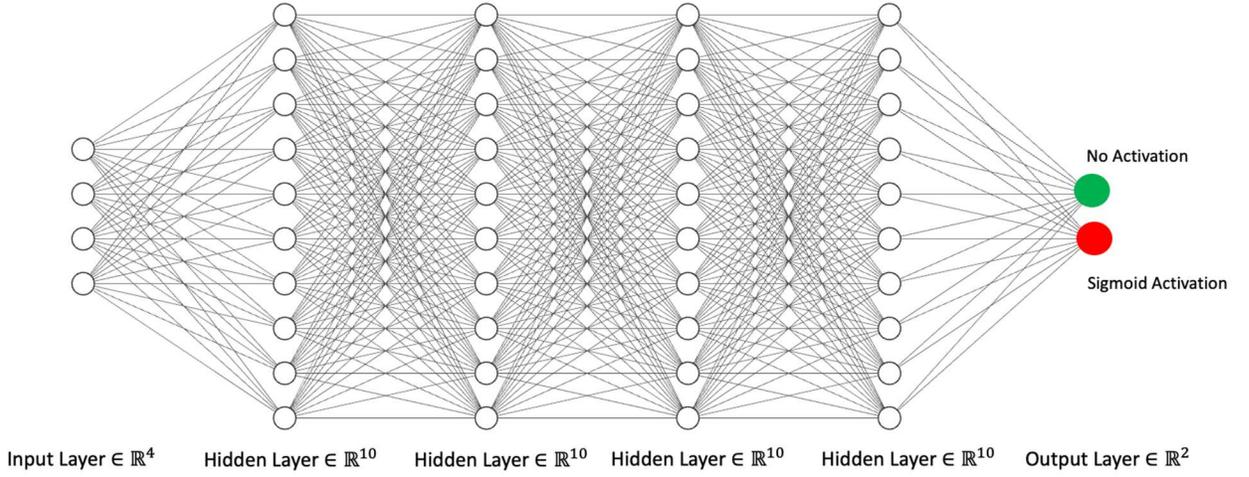

Fig. 3: Binary composition model. An additional binary feature, composition identifier, is fed in the network.

### 4.3 Evaluation Metrics

The metrics employed for evaluating model performance were the mean squared error (MSE) and coefficient of determination ($R^2$). MSE is a commonly used metric, also a loss function in training neural networks for regression problems. It indicates the average squared deviation between the predicted value and the true value. A lower mean squared error indicates that the model makes better predictions.

$$MSE = \frac{\sum_{i=1}^{N}(y_{pred} - y_{true})^2}{N} \qquad (4.3)$$

Equation 4.3 shows the definition of MSE, where $N$ is the sample size, $y_{pred}$ is the predicted value from the model and $y_{true}$ is the true label value of the given sample.

Another metric used in this work was the coefficient of determination, $R^2$, which depicts the extent of the variance in the dependent variable that can be anticiapted from the independent variables. The $R^2$ is defined as

$$R^2 = 1 - \frac{SS_{res}}{SS_{tot}} \qquad (4.4)$$

where, the residual sum of squares is $SS_{res} = \sum_{i=1}^{N}(y_{true} - y_{pred})^2$, the total sum of squares is $SS_{tot} = \sum_{i=1}^{N}(y_{true} - \bar{y}_{true})^2$, and $\bar{y}_{true}$ is the average value of the true values. The values of $R^2$ is usually between 0 and 1. If we have a perfect model that can explain all the variance presented in the data, we will have $R^2$ score of 1.

## 4.4 Experimental Data and Data Setup for Learning

The experimental data used in this study for the binary systems of n-decane/nitrogen ($C_{10}/N_2$) and n-dodecane/nitrogen ($C_{12}/N_2$) were taken from the works of Garcia-Sanchez et al. [6] and Garcia-Cordova et al. [17] respectively. The critical temperatures and pressures for nitrogen, n-decane and n-dodecane are [126.19, 617.8, 658.2] K and [33.97, 21.1, 18] bar respectively. For each of the two mixtures, the machine learning models were trained on isotherm data which lied between the critical temperatures of the individual components. A summary of the data set size used for each isotherm, for the two mixtures, is presented in Table 1.

**Table 1: Summary of VLE Experimental Data Set**

| Mixture | Temperature [K] | No. of points | Source |
|---|---|---|---|
| $C_{10}/N_2$ | 344.6 | 23 | Garcia-Sanchez et al. [6] |
| | 377.4 | 22 | |
| | 410.9 | 23 | |
| | 463.7 | 23 | |
| | 503 | 20 | |
| | 533.5 | 17 | |
| | 563.1 | 14 | |
| $C_{12}/N_2$ | 344.4 | 25 | Garcia-Cordova et al. [17] |
| | 410.7 | 27 | |
| | 463.9 | 27 | |
| | 503.4 | 27 | |
| | 532.9 | 25 | |
| | 562.1 | 23 | |
| | 593.5 | 15 | |

To examine model performance on different training and validation sets, the training of the networks was done via 10-fold cross-validation. In this process, the original data were randomly split into 10 subsets with equal sizes. For each training cycle, nine of the subsets were used for training and the rest one for validation. A complete 10-fold cross validation of the networks consisted of 10 such training cycles where each subset was utilized for validation purpose.

## 5. Results and Discussion

### 5.1 Effect of the Binary Interaction Parameter

Mixing rules which are used along with the EOSs, for multi-component mixtures, require binary interaction parameters ($k_{ij}$) to be able to predict VLE. The inaccurate estimation of such interaction parameters may lead to large discrepancies from the actual equilibrium behavior. Although in this section results for the binary mixture of $C_{10}/N_2$ are discussed, similar observations were also applicable for the binary system of $C_{12}/N_2$. Fig. 4 shows the effect of zero and non-zero $k_{ij}$ on PR-EOS modeled VLE for the binary system of $C_{10}/N_2$, and its comparison against experimental data [6]. The non-zero $k_{ij}$ value of 0.1597 for the binary system of $C_{10}/N_2$ has been taken from the works of Garcia-Sanchez et al. [6], based on minimizing the sum of squared deviations of pressures and mole fraction of phase equilibrium compositions, with corresponding experimental values. The plots consist of mixture equilibrium pressure on the y-axis and $N_2$ liquid ($x_1$) and vapor ($y_1$) mole fractions on the x-axis. The average absolute error (AAE) for the PR-EOS models, calculated according to equation 5.1, is shown in Table 2.

$$AAE = \frac{1}{N}\sum_{i=1}^{N} \frac{|P_{Exp} - P_{model}|}{P_{Exp}} \times 100 \qquad (5.1)$$

where, $N$ is the number of data points, $P_{Exp}$ is the equilibrium pressure from experimental data and $P_{model}$ is the equilibrium pressure estimated either using PR-EOS model or the data driven learning models.

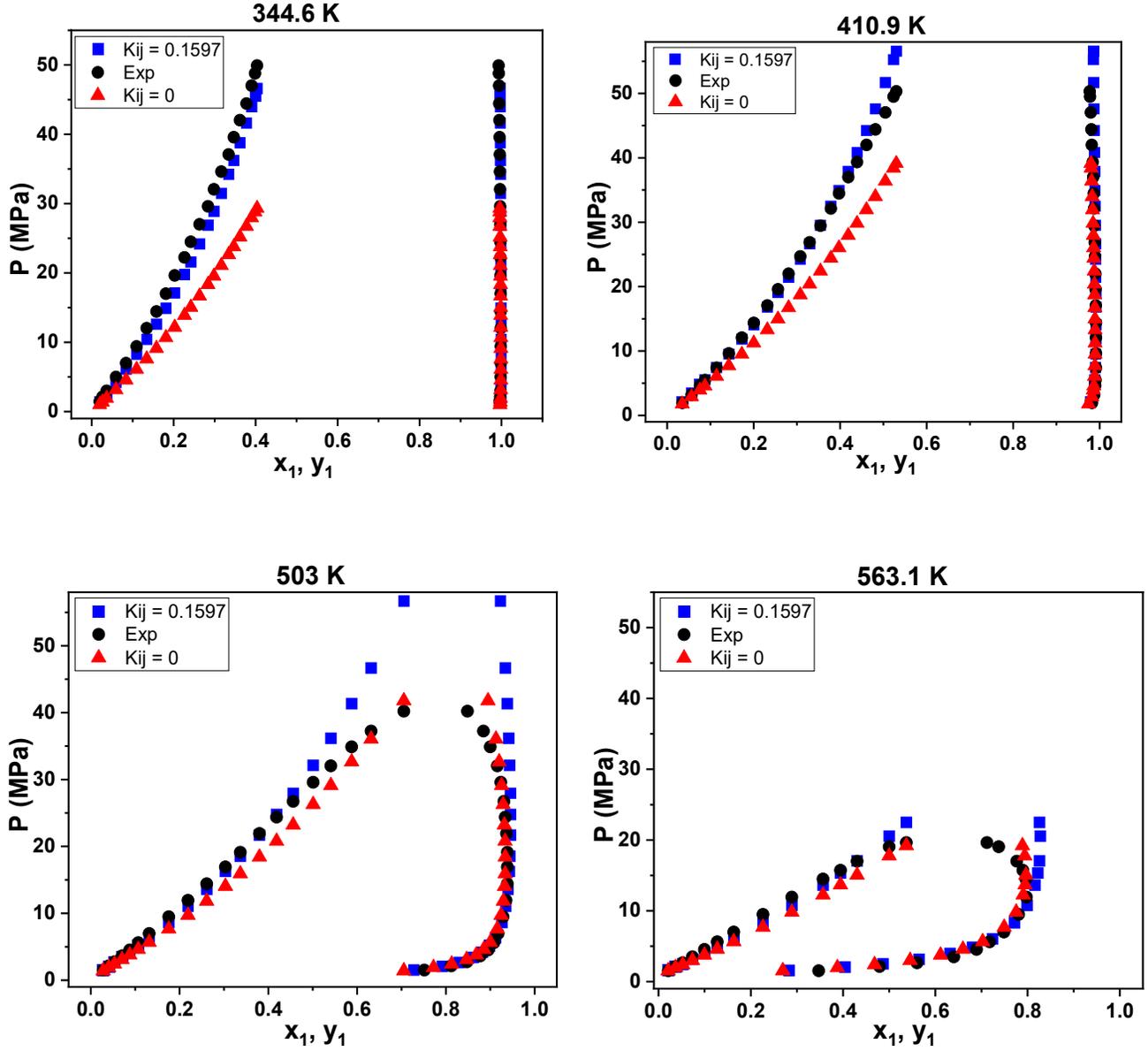

Fig. 4: Effect of binary interaction parameter on VLE for the binary mixture of $C_{10}/N_2$ – a comparison between experimental data (black circles) [6], PR-EOS with $k_{ij} = 0.1597$ (blue squares) and PR-EOS with $k_{ij} = 0$ (red triangles)

Table 2: Absolute Average Error in Pressure Estimation

| Temperature [K] | PR-EOS with $k_{ij} = 0$ | PR-EOS with $k_{ij} = 0.1597$ |
|---|---|---|
| 344.6 | 37.96 % | 10.66 % |
| 410.9 | 21.63 % | 3.42 % |
| 503 | 14.06 % | 7.55 % |
| 563.1 | 14.14 % | 8.1 % |

Fig. 4 and Table 2 show that the PR-EOS model with $k_{ij} = 0.1597$ was able to predict the VLE of the binary system with reasonable accuracy and having AAE values of ~10% or lesser. On the other hand, the performance of the PR-EOS model with $k_{ij} = 0$, improved as the temperature increased from 344.6 K to 563.1 K, with a tendency of underpredicting the equilibrium pressure with large errors of ~38% at low temperatures of 344.6 K. In propulsion applications, high temperature and pressures are more relevant, and under such conditions both PR-EOS models seemed to have the comparable performance for the most part of the VLE curves. However, a closer look reveals that for high temperatures, both PR-EOS models failed to trace the curvatures at the top of VLE curves, as the mixture approached the critical pressure for a given isotherm. In the successive sections, PR-EOS model refers to VLE modeled using PR-EOS and a non-zero binary interaction parameter, estimated by minimizing the sum of squared deviations of pressures and mole fraction of phase equilibrium compositions with corresponding experimental values. For the binary mixture of $C_{12}/N_2$, the $k_{ij}$ value of 0.1561 has been taken from the works of Garcia-Cordova et al. [17].

## 5.2  Unified Data-Driven Model using Interpolation (DD-Int)

In this section, the performance of the unified binary composition model trained using the interpolated dataset is discussed. The shape of the VLE curves for the binary mixtures of $C_{10}/N_2$ and $C_{12}/N_2$ change drastically as temperature rises. Low temperatures are characterized by linear and steep slopes, indicating very high critical pressures for the concerned isotherm. On the other hand, at high temperatures, the liquid and vapor limbs of VLE curve towards each other in an apparent attempt to form a closed loop at low pressures. This also indicates that at higher temperatures, the mixture critical pressure will be lower. Given the change in the VLE behavior as a function of temperature, a data-driven model found it hard to exhibit good performance across all temperatures if it was trained on an equal number of points. As a result, the DD-Int model was trained asymmetrically, with the interpolation dataset having 2000 points for low-temperature range and 3000 points for higher isotherms ($\geq$ 503 K). Figures 5 and 6 show the VLE data on a pressure–composition plot for the binary mixtures of $C_{10}/N_2$ and $C_{12}/N_2$, respectively. Performance of the DD-Int model was gauged against the experimental data and PR-EOS modeled VLE data. From the VLE of both the n-alkane/nitrogen binary mixtures, it can be seen that for a fixed temperature, the solubility of nitrogen increases in the n-alkane rich liquid phase with the increase of pressure. For a fixed pressure, the solubility of nitrogen in the n-alkane rich liquid phase is also found to be directly proportional to the mixture temperature [6,17]. For isotherms above 463 K, the PR-EOS model deviated from the experimental data points at higher pressures. The DD-Int model, on the other hand, was able to capture the change in curvature of the VLE curves at higher pressures for both binary mixtures. For the 563.1 K isotherm of $C_{10}/N_2$ in Fig. 5, the DD-Int model predictions were closer to the PR-EOS modeled data than to the experimental data but still exhibited a change of curvature in the vapor arm of the curve at higher pressures, which was not captured by the EOS model. For the $C_{12}/N_2$ mixture, the DD-Int model showed good agreement with experimental trends across all isotherms and pressure ranges.

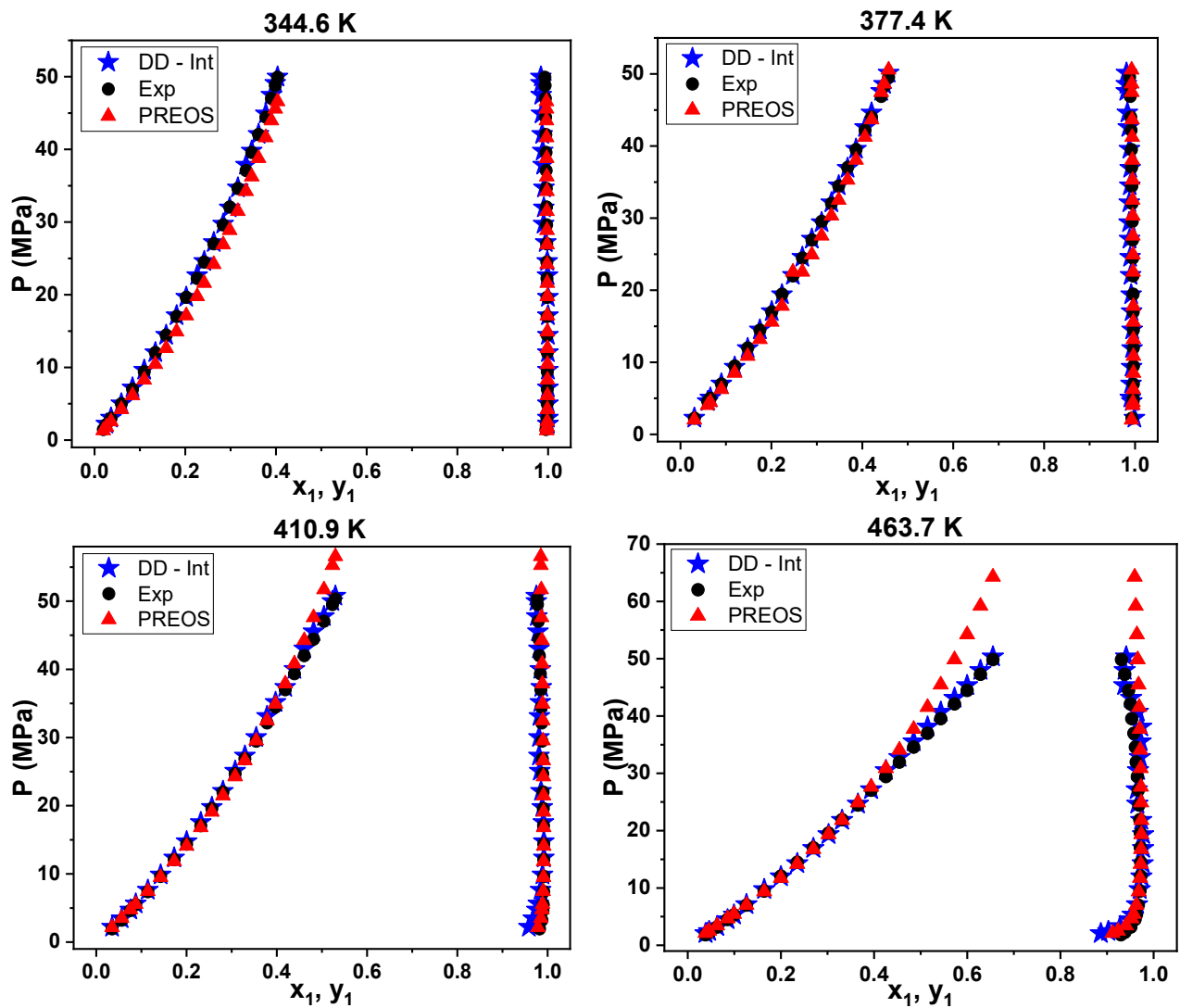

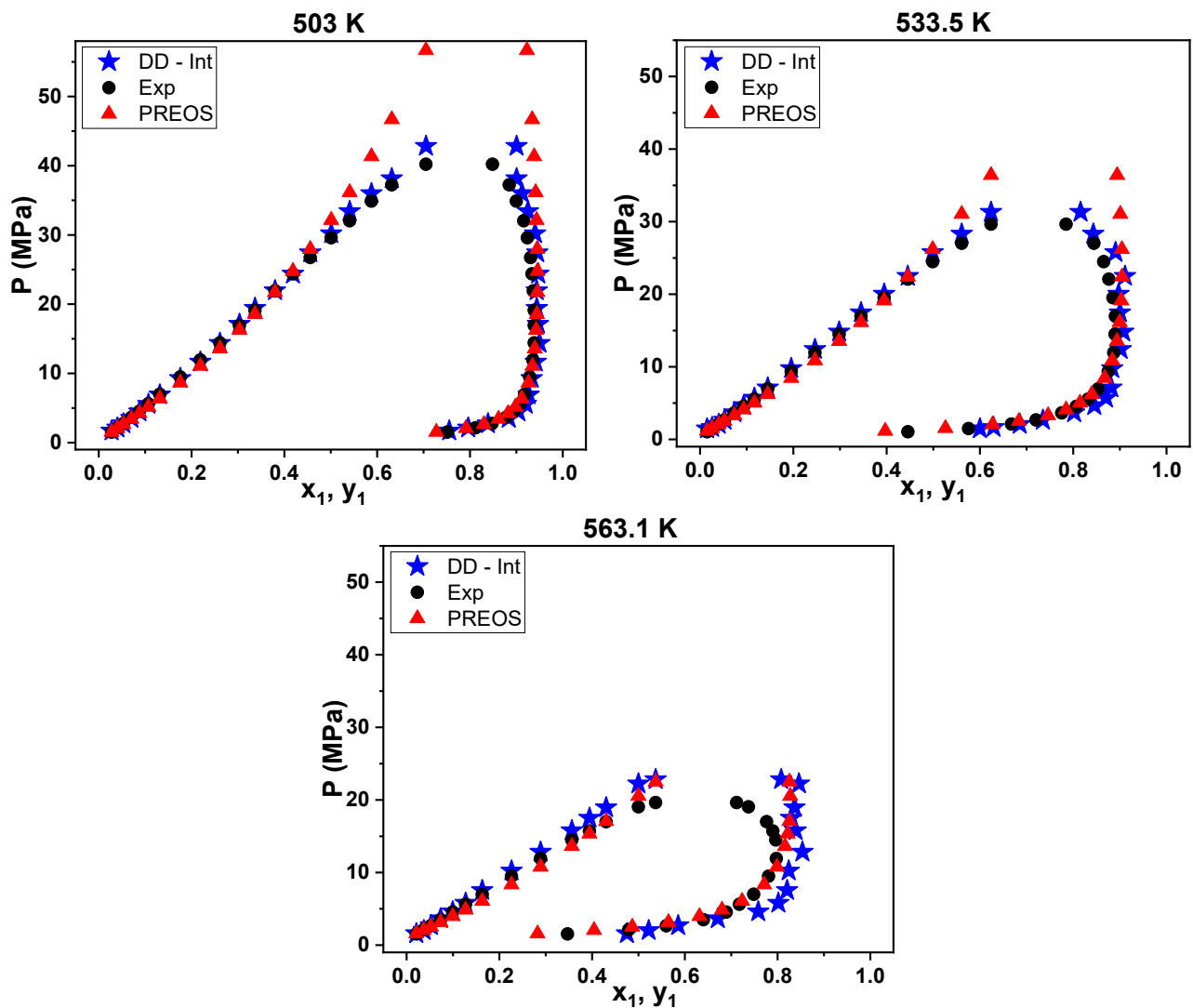

Fig 5. Comparison of VLE for the binary mixture of $C_{10}/N_2$ using: (a) DD-Int model (blue star), (b) Experiment data [6] (black circle) and (c) Peng-Robinson EOS (red triangle)

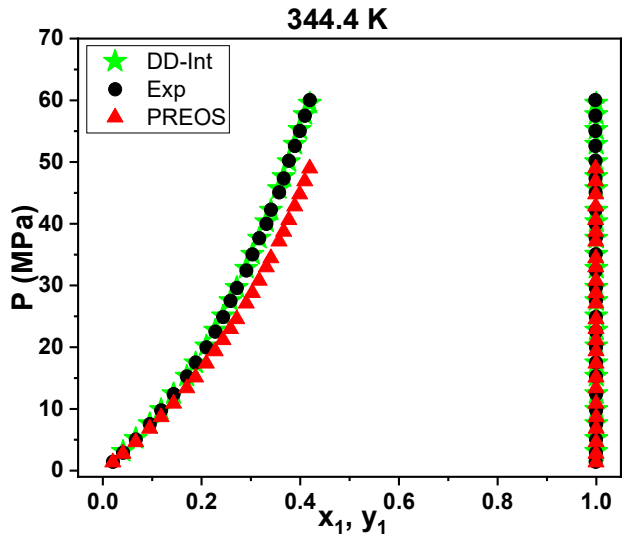
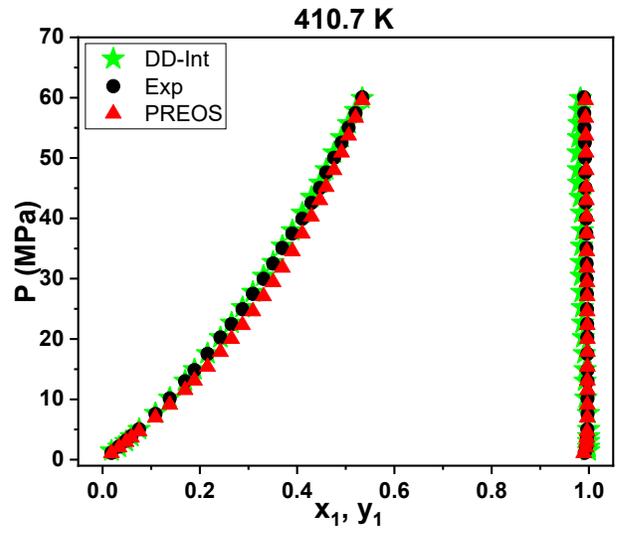
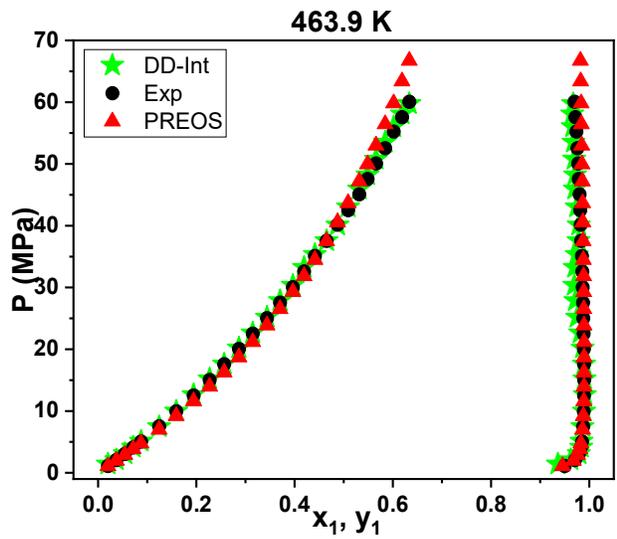
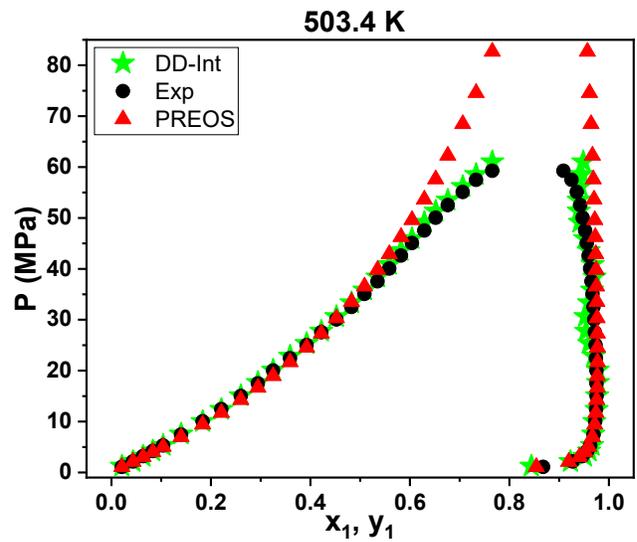

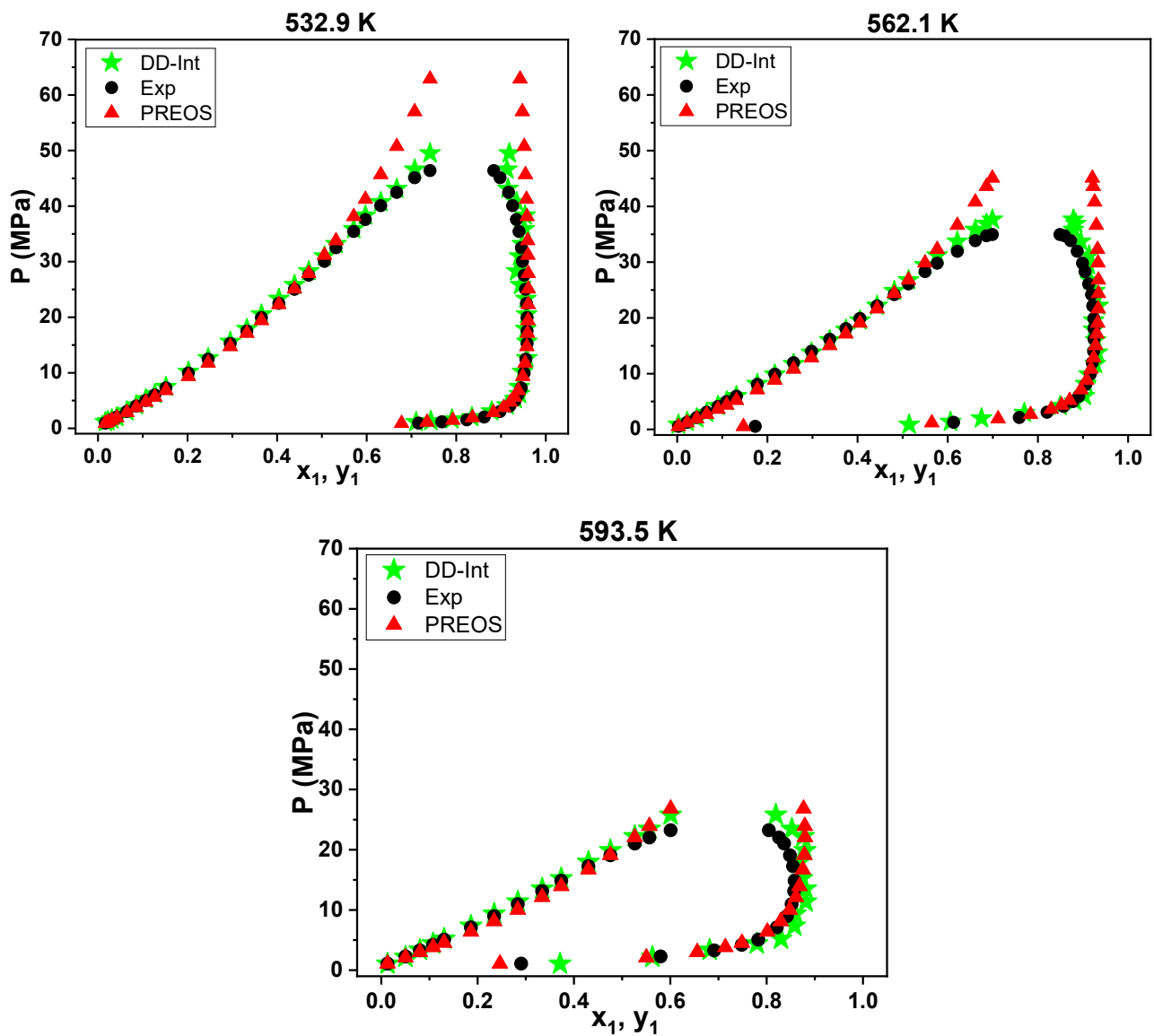

Fig 6. Comparison of VLE for the binary mixture of $C_{12}/N_2$ using: (a) DD-Int model (green star), (b) Experiment data [17] (black circle) and (c) Peng-Robinson EOS (red triangle)

Fig. 7 shows the mean value of the absolute percentage errors in pressure estimation as a function of temperature, for the EOS and DD-Int model. For both mixtures, it can be seen that the machine learning-based model had much lower error as compared to the PR-EOS model. For the binary mixture of $C_{10}/N_2$, using the DD-Int model, the errors in pressure estimation were less than 3.4%, except for the 563.1 K, where the mean error jumped to ~7%, but still lower than the EOS model error. The error in estimating the vapor phase composition was also less than 3%, with the exception of ~7.5% for the 563.1 K isotherm. DD-Int modeled VLE for the binary system of $C_{12}/N_2$ had pressure errors in the range of [0.7 – 3.8] %, which was considerably lower compared to the error from the PR-EOS model. The pressure error from the DD-Int model exhibited a gradual increase in magnitude with the rise of temperature. The $C_{12}/N_2$ vapor phase composition predicted using the DD-Int model had errors in the range of [0.1 – 3.2] %, as opposed to [5 - 15] % from the PR-EOS model.

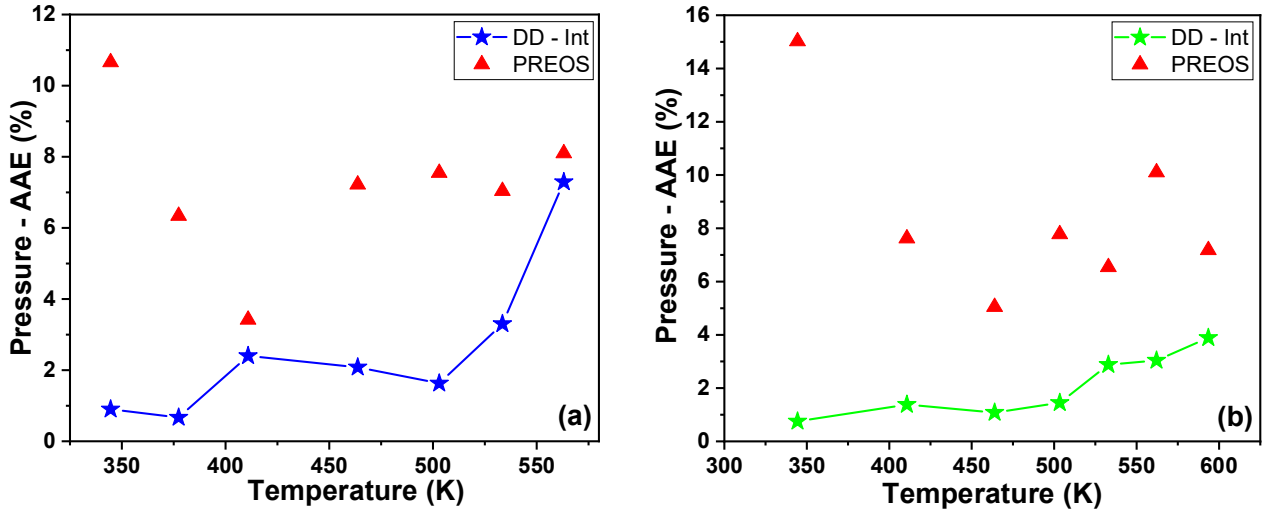

Fig 7. Model performance in estimating mixture equilibrium pressure: data driven interpolation model (DD-Int) vs PR-EOS for the binary mixtures of (a) $C_{10}/N_2$ and (b) $C_{12}/N_2$

## 5.3 Unified Data-Driven Model using Individual Temperature Trained Models (DD-Ind-T)

In this section, the unified data-driven model (DD-Ind-T) were trained using the individually trained temperature models. The individual models were trained using the limited experimental data set available. The average model performance metric for each of the individual isotherm models are given in Table 3.

Table 3: $R^2$ and MSE for the individual temperature models

| Mixture | Temperature [K] | MSE | $R^2$ |
|---|---|---|---|
| $C_{10}/N_2$ | 344.6 | 0.0536 | 0.9996 |
|  | 377.4 | 0.0649 | 0.9997 |
|  | 410.9 | 0.0290 | 0.9992 |
|  | 463.7 | 0.0364 | 0.9998 |

|  | 503 | 0.1349 | 0.9944 |
|---|---|---|---|
|  | 533.5 | 0.0378 | 0.9995 |
|  | 563.1 | 0.0893 | 0.9950 |
| $C_{12}/N_2$ | 344.4 | 1.2012 | 0.9924 |
|  | 410.7 | 0.6130 | 0.9970 |
|  | 463.9 | 1.0047 | 0.9661 |
|  | 503.4 | 0.5138 | 0.9928 |
|  | 532.9 | 0.4469 | 0.9969 |
|  | 562.1 | 0.0731 | 0.9987 |
|  | 593.5 | 0.0491 | 0.9680 |

The individually trained models were used to generate 2000 data points for each temperature by taking randomly sampled liquid phase mixture compositions ($x_1$ and $x_2$). This newly generated dataset was then used to train the new unified data-driven model (DD-Ind-T). Figures 8 and 9 show the DD-Ind-T modeled VLE data on a pressure – composition plot for both the binary mixtures, and their comparison with experimental data and PR-EOS model. Similar to the DD-Int model, the DD-Ind-T model was also capable of capturing the correct trend of increase in nitrogen solubility in the n-alkane rich liquid phase, with the increase of pressure and temperature [6,17].

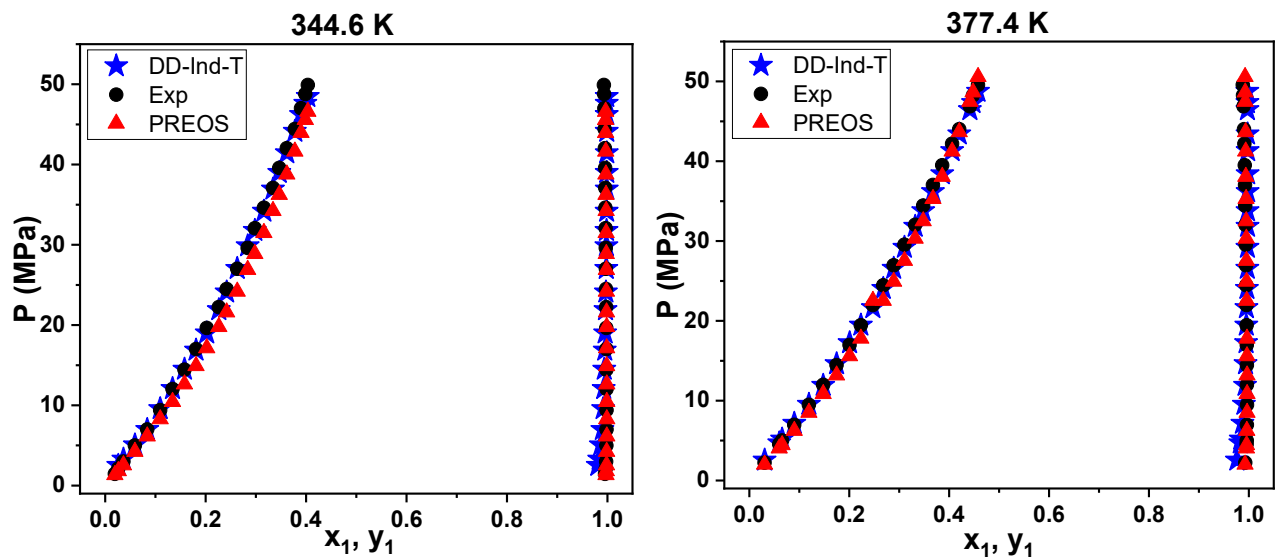

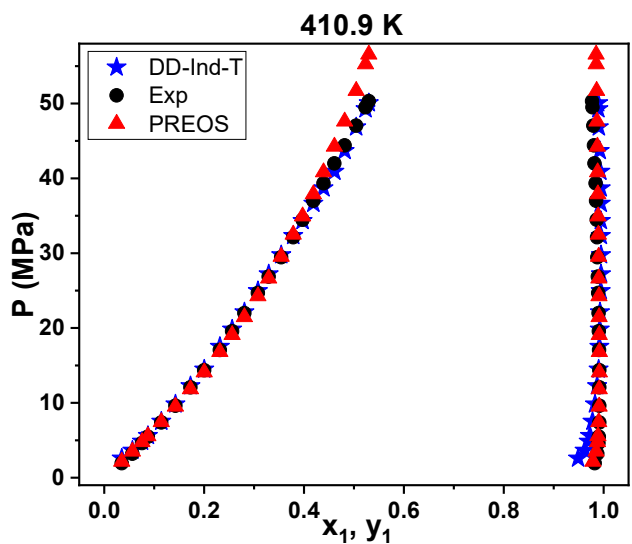
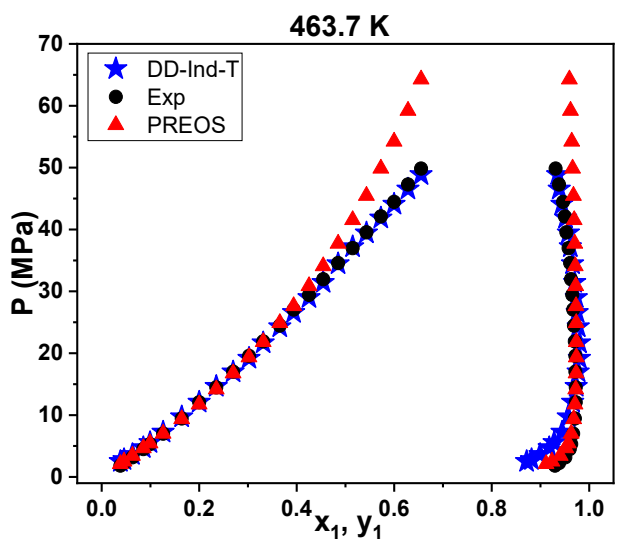
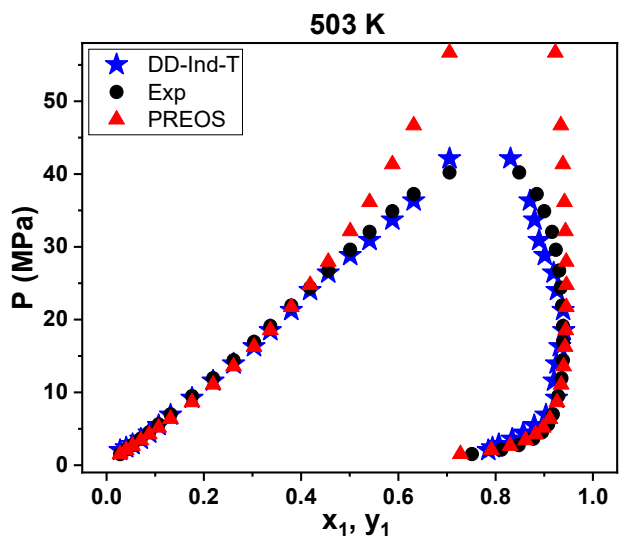
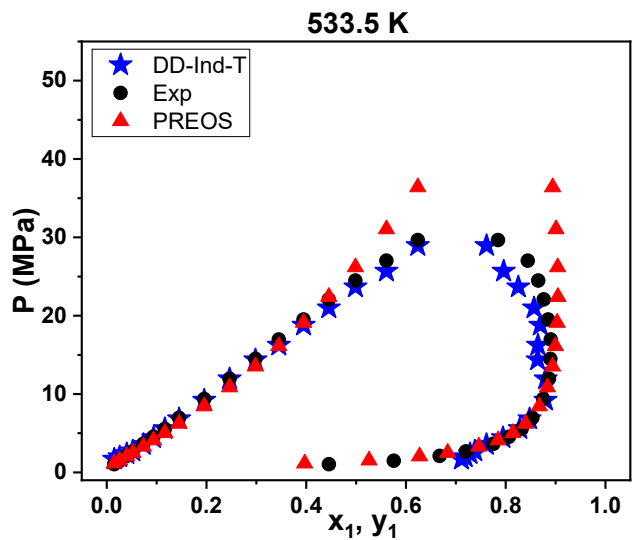
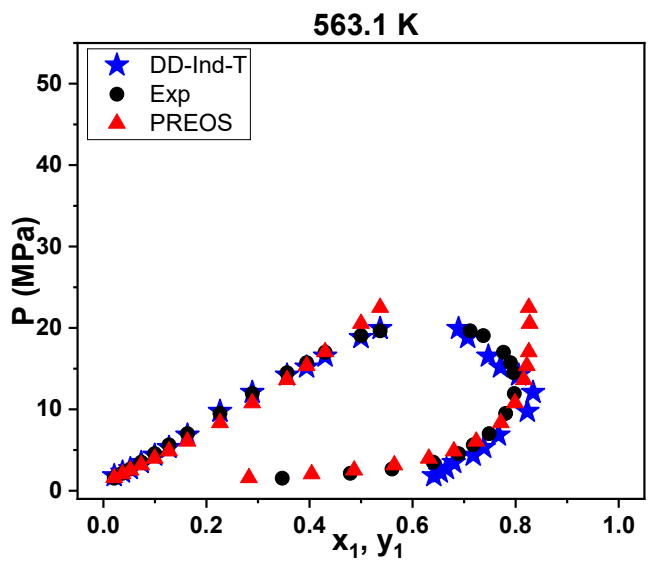

Fig 8. Comparison of VLE for the binary mixture of $C_{10}/N_2$ using: (a) DD-Ind-T model (blue star), (b) Experiment data [6] (black circle) and (c) Peng-Robinson EOS (red triangle)

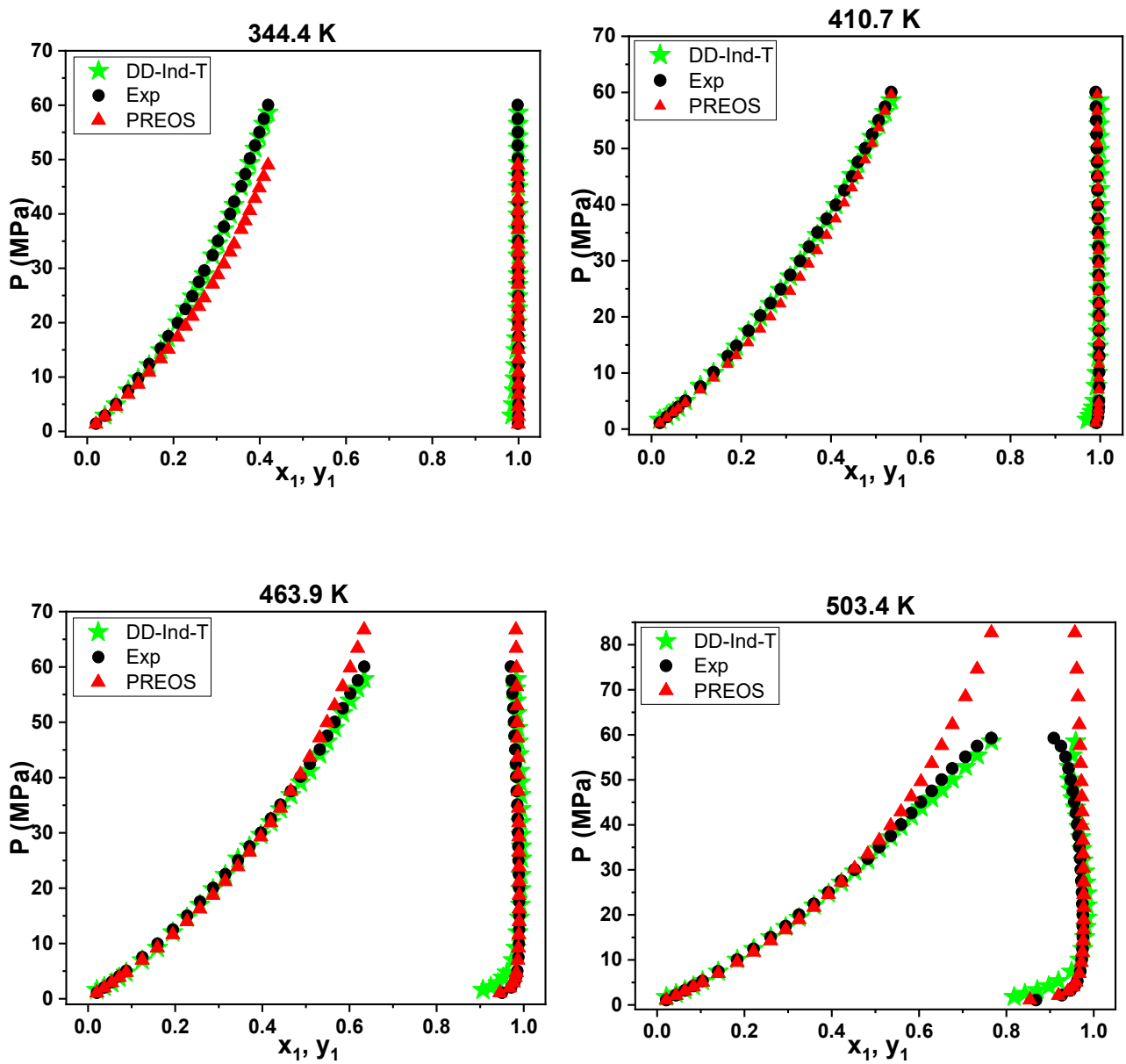

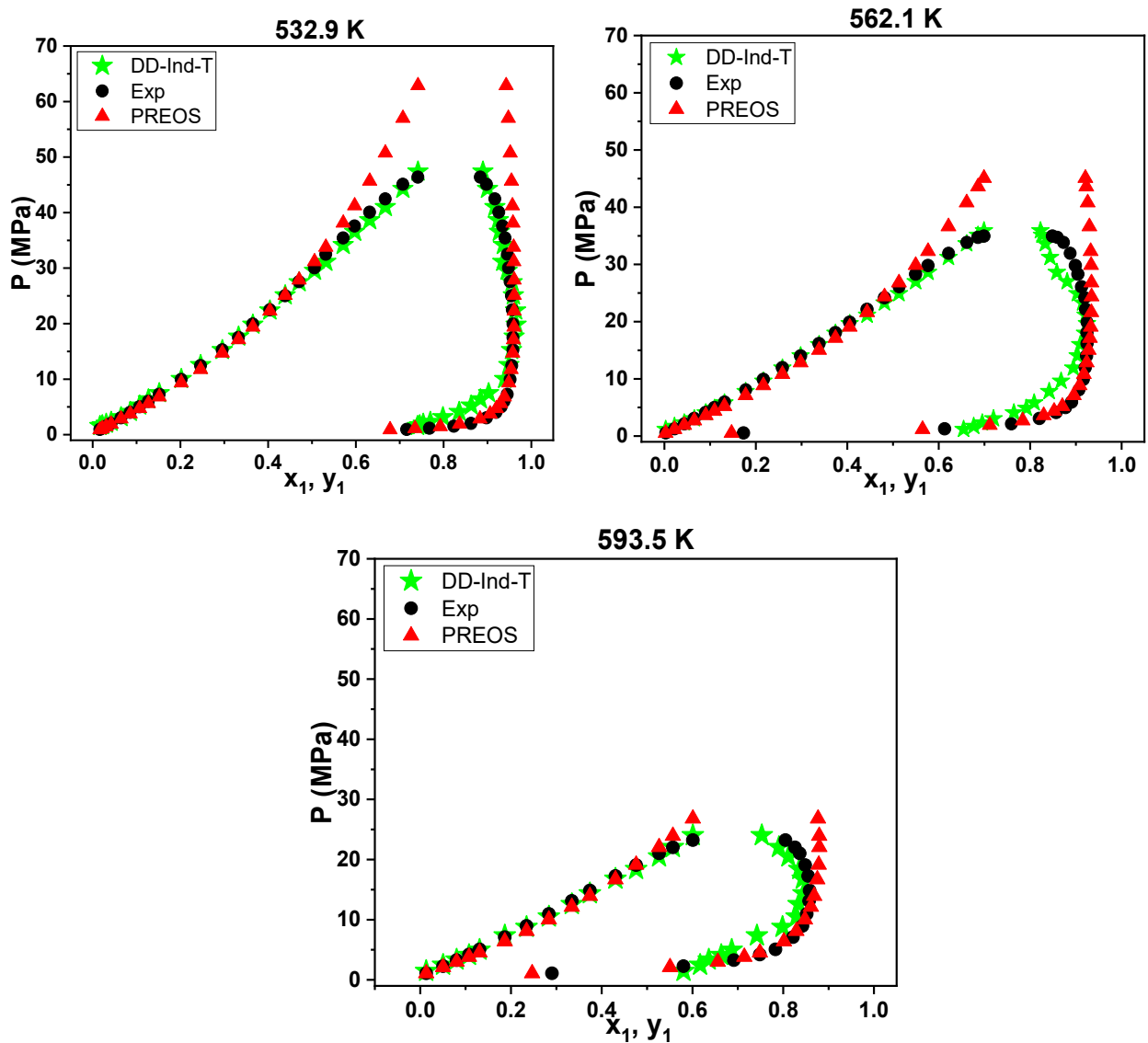

Fig 9. Comparison of VLE for the binary mixture of $C_{12}/N_2$ using: (a) DD-Ind-T model (green star), (b) Experiment data [17] (black circle) and (c) Peng-Robinson EOS (red triangle)

The DD-Ind-T model showed improvement over the DD-Int model in being able to track the changing curvatures of the vapor phase curve at high pressures, even for the high temperature (563.1 K) isotherm of the $C_{10}/N_2$ system. A comparison of the average absolute error plots in predicting pressure, using the DD-Ind-T and PR-EOS models, for the binary mixtures of $C_{10}/N_2$ and $C_{12}/N_2$ is shown in Fig. 10. For the $C_{10}/N_2$ mixture, the pressure errors were in the range of [1.5 – 5.2] %, whereas the PR-EOS model had errors of the order of [3.4 – 10.6] %. The DD-Ind-T model also exhibited a slight increase in the mean error magnitudes with the increase of temperature. The pressure and vapor composition errors, for $C_{12}/N_2$, using the DD-Ind-T model, were in the ranges of [1.3 – 4.2] % and [0.5 – 5.3] % respectively. Using the PREOS model resulted in higher errors for pressure of the order of [5 – 10.1] % for the $C_{12}/N_2$ binary system. However, the DD-Ind-T model found it hard to correctly estimate the vapor phase composition at extremely low pressures of 1 MPa for higher temperature isotherms (T ≥ 533 K).

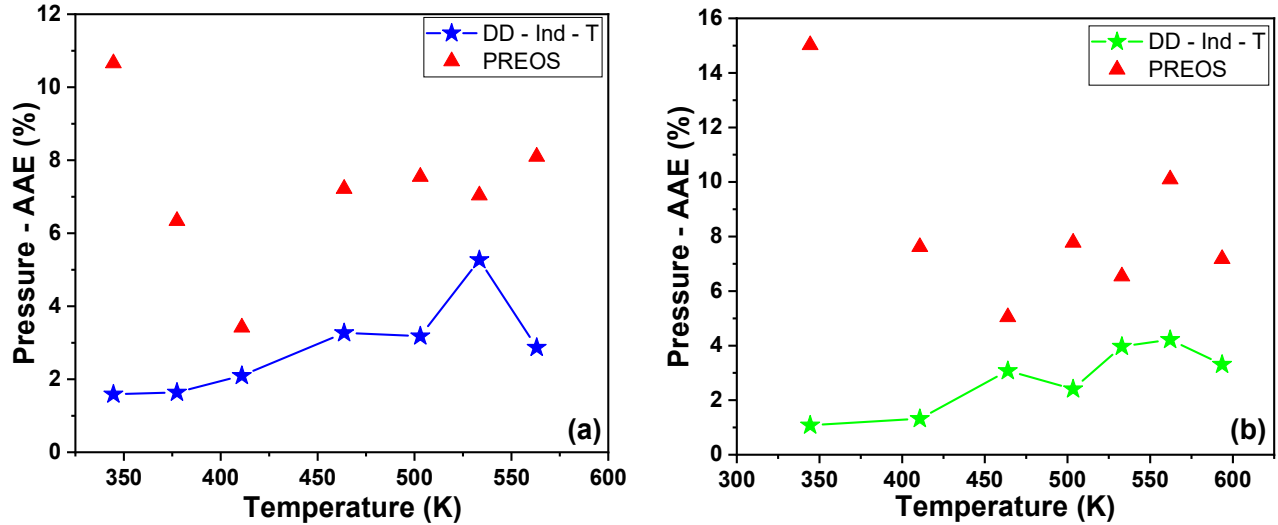

Fig 10. Model performance in estimating mixture equilibrium pressure: data driven individual temperature trained model (DD-Ind-T) vs PREOS for the binary mixtures of (a) $C_{10}/N_2$ and (b) $C_{12}/N_2$

## 6. Conclusions

Data–driven learning-based unified models were proposed in this study to predict the vapor-liquid equilibrium of n-decane/nitrogen and n-dodecane/nitrogen binary mixtures. One of the models (DD-Int) was trained based on the interpolated dataset generated using the experimental data. The second model (DD-Ind-T) was trained based on the data generated from individually trained temperature based sub-models. VLE of the two binary mixtures was also modeled using the Peng-Robinson equation of state and compared against that of the machine learning-based models.

Both machine learning-based models were able to predict the equilibrium pressure of the binary mixtures with considerably less error as compared to the PR-EOS model. The improvement in accuracy came with the added advantage of being completely independent from a binary interaction parameter. This parameter plays a crucial role in the mixing rules used in the EOS based VLE model but is not readily available due to the lack of properties/experimental data of the pure components, which are not readily available.

The two machine learning-based models exhibited an increase in error in estimating equilibrium pressure, as the temperature of the mixture increases. The PR-EOS model failed to trace the curvature of the vapor phase exhibited at high pressures, close to the mixture critical point, at large temperatures. Both DD-Int and DD-Ind-T models, on the other hand, were successful in tracing the change of slopes of the vapor phase curves at higher pressures and temperatures. The DD-Ind-T model however was not able to correctly predict the vapor phase composition at low pressures (~ 1 MPa) and high temperatures.

The availability of experimental data on VLE of long chained n-alkane and nitrogen mixtures at high pressures and temperatures were few and far between. Both the models proposed in this study were shown to be robust, even though very limited experimental data points were used (each isotherm having 15–27 data points). The results suggested that the DD-Int and DD-Ind-T models could be used as a non-iterative, computationally efficient tool to predict VLE data for hydrocarbon mixtures such as $C_{10}/N_2$ and $C_{12}/N_2$ mixtures at high pressures and temperatures, even close to mixture critical point. IN addition, the results also suggested that such models can be used in the future to help determine parameters in EOS models such as the temperature-dependent binary interaction parameter for given mixtures.

**Acknowledgments**

GL and YS gratefully acknowledge the support from the National Science Foundation (DMS-1555072, DMS-1736364, CMMI-1634832, and CMMI-1560834), and Brookhaven National Laboratory Subcontract 382247, ARO/MURI grant W911NF-15-1-0562, and U.S. Department of Energy (DOE) Office of Science Advanced Scientific Computing Research program DE-SC0021142.